\documentclass[journal=apchd5,manuscript=article]{achemso}

\usepackage[version=3]{mhchem} 
\usepackage{orcidlink}
\usepackage{hyperref}
\usepackage{lineno}
\SectionNumbersOff
\usepackage{array}
\usepackage{amsmath, amssymb}
\usepackage{tabularx} 
\usepackage{graphicx} 
\usepackage{cancel}
\usepackage[normalem]{ulem}

\usepackage{xcolor}

\hypersetup{
  colorlinks=false,      
  linkcolor=black,      
  citecolor=black,
  urlcolor=black
}

\definecolor{figcolor}{RGB}{0,102,204}   
\definecolor{tabcolor}{RGB}{0,153,0}     
\definecolor{algcolor}{RGB}{204,0,102}   

\makeatletter
\newenvironment{funding}{%
\acs@section*{\fundingname}%
}{}
\newcommand*\fundingname{Funding}
\makeatother

\makeatletter
\newenvironment{datastatement}{%
\acs@section*{\datastatementname}%
}{}
\newcommand*\datastatementname{Data and Code Availability Statement}
\makeatother

\makeatletter
\newenvironment{notes}{%
\acs@section*{\notename}%
}{}
\newcommand*\notename{Notes}
\makeatother

\renewcommand{\tabularxcolumn}[1]{m{#1}} 

\author{Arpan Sur\,\orcidlink{0009-0003-5833-9998}}
\affiliation[BUET]
{Department of Electrical and Electronic Engineering, Bangladesh University of Engineering and Technology, Dhaka, Bangladesh}
\altaffiliation{These authors contributed equally to this work. }

\author{Sudipta Saha\,\orcidlink{0009-0003-9110-8059}}
\affiliation[BUET]
{Department of Electrical and Electronic Engineering, Bangladesh University of Engineering and Technology, Dhaka, Bangladesh}
\altaffiliation{These authors contributed equally to this work. }

\email{sudiptasaha@ari.buet.ac.bd}

\author{Chih-Yu Lee\,\orcidlink{0000-0001-7437-7343}}

\affiliation[UMD]
{Department of Materials Science and Engineering, University of Maryland, College Park, MD, USA}

\author{Ichiro Takeuchi\,\orcidlink{0000-0003-2625-0553}}
\affiliation[UMD]
{Department of Materials Science and Engineering, University of Maryland, College Park, MD, USA}
\alsoaffiliation[3]
{Maryland Quantum Materials Center, University of Maryland, College Park, MD, USA}
\email{takeuchi@umd.edu}

\title {Segmentation-Engineered Ge\textsubscript{4}Sb\textsubscript{6}Te\textsubscript{7} Switch on SOI platform for Multilevel Non-volatile Photonic Neural Inference}


\begin{document}

\begin{tocentry}
\includegraphics[width=3.25in,height=1.75in]{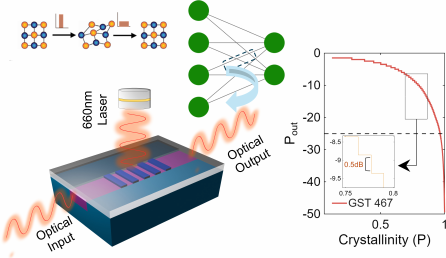}




\end{tocentry}
\begin{abstract}
Phase-change materials (PCMs) have emerged as key enablers of non-volatile, ultra-compact photonic switches for energy-efficient deep neural network (DNN) applications. In this work, we computationally investigate the recently discovered $\mathrm{Ge_{4}Sb_{6}Te_{7}}$ (GST-467) as a high-contrast optical PCM and demonstrate its suitability for multilevel photonic computing. We experimentally measure the complex refractive indices of amorphous and crystalline GST-467 and use them to numerically propose a segmented silicon-on-insulator photonic switch optimized at 1550 nm. Three-dimensional finite-difference time-domain simulations show that segmentation significantly enhances the extinction ratio while maintaining low insertion loss, yielding a design figure of merit more than sevenfold higher than that of an unsegmented design. Multiphysics thermo-optical simulations further demonstrate efficient, reversible laser-induced switching with sub-nJ energy requirements for crystallization and amorphization. Compared with established GST, GSST, and GSS compositions, GST-467 provides the largest transmittance contrast and supports up to 48 resolvable optical states. When deployed as multilevel weights in photonic DNN architectures, the GST-467 switch achieves superior classification accuracy on EMNIST and Fashion-MNIST benchmarks. These results position GST-467 as a highly promising PCM for scalable, low-energy photonic computing and neuromorphic hardware.
\end{abstract}

\section{Introduction}
The rapid advancement of artificial intelligence (AI) and neural networks has significantly increased demand for high-performance, energy-efficient computing platforms \cite{wang2025integrated,ning2024photonic}. As neural networks grow larger and more complex, traditional computing architectures face severe limitations in bandwidth, latency, and energy consumption. These constraints primarily arise from the frequent data transfers between memory and processing units, which dominate both performance and power budgets in modern AI systems. The conventional von Neumann architecture, which separates memory and computation, has become a critical bottleneck for AI workloads \cite{sebastian2018tutorial}. The repeated exchange of data between the processor and memory leads to the von Neumann bottleneck, which causes inefficient resource use and limits scalability \cite{sebastian2020memory,lian2022photonic,li2025photonics}. To overcome these challenges, in-memory computing has emerged as a promising method that enables computation directly within memory arrays, thereby minimizing data transfer and achieving substantial improvements in speed and energy efficiency \cite{xia2019memristive,ding2019phase,sebastian2018tutorial}. In these architectures, the same physical device is used to store data and perform computation. This approach enables vector-matrix multiplications and accumulation operations to be performed locally, resulting in high parallelism and reduced latency. However, several device-level challenges remain, such as conductance drift, programming variability, and noise, which can affect computational precision \cite{gong2018signal,chen2025emerging}. Performing computation with light provides a compelling path to mitigate data-movement and interconnect penalties. In fact, photonic integrated circuits (PICs) offer ultrahigh bandwidth, low latency, and immunity to electromagnetic crosstalk, while enabling non-volatile, reconfigurable elements on silicon photonics when combined with suitable materials \cite{shekhar2024roadmapping,xiao2023recent}. By co-locating storage and computation optically, photonic platforms can reduce the need for repeated electro-optic conversions and avoid the static power associated with volatile tuning mechanisms, thereby improving scalability for deep neural network inference (DNN) \cite{chen2025emerging}.

Chalcogenide phase-change materials (PCMs) have gained increasing prominence in PICs and non-volatile memory applications due to their reversible modulation of optical and electrical properties between amorphous and crystalline states. Traditional options such as $\mathrm{VO_{2}}$ are volatile, which requires standby power, whereas $\mathrm{Ge_{2}Sb_{2}Te_{5}}$ (GST-225) is non-volatile and has been widely used \cite{chen2023review,wu2011ultrafast,loke2012breaking}. Tellurium-based phase-change materials, namely the GeTe, GeSbTe (GST), $\mathrm{GeSbSeTe}$ (GSST), and SbTe families, predominantly enable amplitude modulation due to their large refractive-index contrast. Recently, low-loss optical PCMs such as $\mathrm{Ge_{2}Sb_{2}Se_{4}Te_{1}}$ (GSST-2241), $\mathrm{Sb_{2}S_{3}}$, and $\mathrm{Sb_{2}Se_{3}}$ have been investigated for designing phase-only photonic switches due to smaller refractive index contrast \cite{zhang2019broadband}. Among the various types of PCM-based photonic switches, all-optical photonic switches offer distinct advantages over their electro-optic or electrically driven counterparts \cite{shen2017deep}. First, they eliminate resistive-capacitive delays and repeated electro-optic conversions, enabling sub-nanosecond state changes and GHz-rate write operations using picosecond optical pulses. Also, readout can be performed rapidly with low-energy probes \cite{saha2023engineering,rios2015integrated}. Second, because the programmed phase states are non-volatile, no static bias is required to maintain the weights, unlike thermo-optic or carrier-based electro-optic tuners, which consume continuous power \cite{zhang2019miniature}. Third, fully optical signal paths exhibit low residual crosstalk and high compatibility with existing optical network architectures \cite{sun2023all,rios2015integrated}. Collectively, these attributes yield superior energy–latency characteristics and better cascaded system behavior for large PICs. A segmented switch topology further amplifies these material advantages.
By distributing multiple PCM segments along a waveguide, one can realize multilevel optical weights with deterministic level placement, reduced device-to-device thermal crosstalk at the array level, and improved calibration stability.
Segmentation has been shown to (i) expand modulation depth and optical weight dynamic range \cite{zhang2024nonvolatile}, (ii) lower insertion and reflection losses, especially in the low-loss (amorphous) state \cite{quan2022nonvolatile}, and (iii) provide smoother field distributions that enable reliable, cumulative programming using identical optical pulses \cite{cheng2017chip}. These properties directly translate into higher effective bit depth per length, improved cascading, and energy efficiency, which are vital for large neural networks.

Recently, a newly discovered phase-change material, $\mathrm{Ge_{4}Sb_{6}Te_{7}}$ (GST-467) \cite{kusne2020fly}, has attracted attention as a promising candidate for optical phase-change memory applications. Relative to GST-225, GST-467 exhibits higher optical contrast between amorphous and crystalline states in the near-infrared, attributed to a larger band-gap difference \cite{kusne2020fly,khan2023energy}. This enhanced contrast increases switching dynamic range and the number of resolvable multilevel states. Previous studies indicate gradual, near-symmetric potentiation and depression using identical electrical pulse schemes, which is advantageous for weight updates with reduced energy and simplified control \cite{wu2024novel,hamid2024low}. GST-467 offers an attractive trade-off between contrast loss and programmable level behavior for scalable photonic computing. Kusne \textit{et al.} \cite{kusne2020fly} demonstrated a GST-467 photonic switching device by sputtering a 30~nm-thick GST-467 film onto a 330~nm-thick $\mathrm{Si_{3}N_{4}}$ layer and patterning 500~nm-diameter disks on top of a 1.2~$\mu$m-wide waveguide. While $\mathrm{Si_{3}N_{4}}$ offers low-loss operation, it typically requires larger device footprints, motivating exploration of more compact implementations for scalable photonic integration. To date, GST-467 switching on the silicon-on-insulator (SOI) platform and the role of the GST-467 segmented structure remain unexplored. In this paper, we address these gaps by investigating segmented GST-467 switches on SOI and demonstrating a sub-micron-footprint device suitable for more compact integrated photonic architectures.

In this work, we present a comprehensive computational investigation of the novel GST-467 phase-change material for optical switching and photonic computing applications. The complex refractive index of GST-467 was determined experimentally by spectroscopic ellipsometry, and the resulting optical constants were then used to numerically design a SOI-based photonic switch operating at 1550 nm. The switching performance was further enhanced by incorporating a segmented PCM top-cladding configuration, which significantly increased the transmittance contrast between the amorphous and crystalline states. Moreover, we numerically analyzed the laser-induced phase-transition dynamics of GST-467 to identify the optimal pulse power and width for efficient switching. Finally, leveraging the transmittance-crystallinity relationship, the designed photonic switch was evaluated for DNN inference tasks using the EMNIST \cite{cohen_afshar_tapson_schaik_2017} and Fashion-MNIST \cite{xiao2017} datasets, and its performance was benchmarked against other state-of-the-art phase-change materials from the GST, GSST, and GSS families.

\begin{figure}[!ht]
\centering\includegraphics[width=\textwidth]{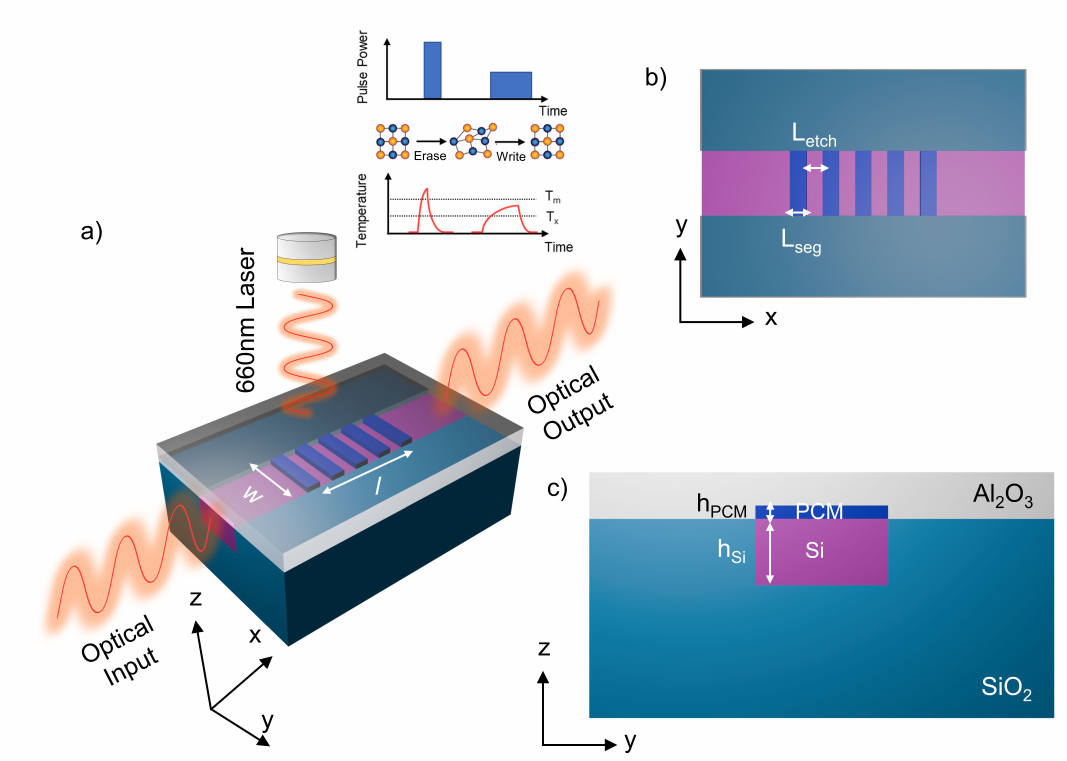}
\caption{(a) 3D schematic representation of the proposed optical switch. The parameters $l$ and $w$ denote the total length and width of the segmented GST sections, respectively. A laser-induced optical phase-switching mechanism is shown for the proposed switch. A focused 660nm laser beam acts as the switching signal, selectively illuminating the GST-467 region. The phase transition between the amorphous and crystalline states is achieved by precisely controlling the laser pulse power and duration. The inset illustrates the bidirectional switching process. The prior transition corresponds to an amorphization process via melt-quenching, induced by a high-intensity, short-duration laser pulse. In contrast, the latter transition denotes recrystallization through controlled thermal annealing, induced by a moderate-power, longer-duration laser pulse. (b) The top view of the switch layout is shown, where $\mathrm{L_{seg}}$ and $\mathrm{L_{etch}}$ represent the lengths of the individual GST-467 and etched sections. (c) Cross-sectional view along the vertical y-z plane of the switch structure. The heights of the GST-467 segment and the silicon waveguide are indicated by h\textsubscript{PCM} and h\textsubscript{Si}, respectively. An Al\textsubscript{2}O\textsubscript{3} layer serves as the cladding for the PCM layer, while SiO\textsubscript{2} serves as the cladding material around the Si core.}
\label{fig:structure}
\end{figure}

\section{Device Design and Methodology}
\subsection{Structure Description and Laser-Induced Phase Change}
\autoref{fig:structure}(a) illustrates the proposed switch structure using GST-467. To ensure compatibility with the mature silicon photonics platform, we adopted the SOI configuration. The crystalline silicon waveguide core has a width ($w$) of 500 nm and a height (h\textsubscript{Si}) of 220 nm, consistent with many silicon foundries \cite{chrostowski2015silicon, xu2014silicon}. The silicon core was placed on a buried oxide (BOX) layer, and Al\textsubscript{2}O\textsubscript{3} was selected as the top cladding oxide, based on other PCM switch structures \cite{yoon2017uv, fang2022ultra}, for its stronger optical confinement, effective diffusion barrier, and surface passivation, as shown in \autoref{fig:structure}(c). The PCM layer was deposited over the silicon core and has a thickness (h\textsubscript{PCM}) of 30 nm \cite{fang2022ultra, sun2025microheater}. Unlike prior works on GST-467-based photonic switches, our proposed structure employs a segmented topology, as depicted in \autoref{fig:structure}(b). The lengths of the PCM and the etched segment are indicated by $\mathrm{L_{seg}}$ and $\mathrm{L_{etch}}$. 
The total device length, $\ell$, shown in \autoref{fig:structure}(a) can be defined as follows: $\ell = \mathrm{N} \times \mathrm{L_{seg}} + (\mathrm{N}-1) \times \mathrm{L_{etch}}$.


Additionally, \autoref{fig:structure}(a) schematically depicts the laser-induced switching mechanism in a GST-467-based photonic platform. This setup employed a focused laser beam as the external stimulus to modulate the local phase state of the GST-467 layer integrated on a dielectric waveguide. Under a short, high-intensity laser pulse, the local temperature of the GST-467 film rapidly exceeded its melting point, $\mathrm{T_m \approx 500-540^\circ C}$ \cite{khan2023energy}. Upon pulse termination, ultrafast quenching prevented atomic rearrangement, resulting in a disordered amorphous phase. This amorphization (erase) operation produced a low refractive index in GST-467, enabling a high-transmission optical state in the device. Conversely, when the amorphous material was irradiated with a low-intensity, longer-duration pulse, the temperature rose to $\mathrm{200-220^\circ C}$, near the crystallization temperature ($\mathrm{T_c}$) \cite{khan2023energy}, facilitating nucleation and grain growth that restored the crystalline order. This crystallization (write) process converted the material back to its high-index phase, reducing transmission and enabling robust, non-volatile optical modulation. Repeating this write-and-erase operation allows bidirectional control of these states, underpinning the reconfigurable operation of GST-467 photonic elements in the proposed switch. 
Here, the focused 660~nm beam covers the entire segmented region and programs all segments into a single, spatially uniform crystallinity, rather than addressing them individually. Thus, each switch stores one multilevel weight, encoded as a crystalline fraction. Segmentation here is an optical mode- and index-engineering strategy that raises the extinction ratio and design figure of merit, and is not a per-segment digital-addressing scheme.

\subsection{Optical Simulation Methodology}
Three-dimensional finite-difference time-domain (FDTD) simulations were performed using Ansys Lumerical FDTD Solutions to analyze the optical transmission and propagation behavior of the proposed GST-467 switch. The spatial mode distributions and wavelength-dependent modal characteristics were determined by solving Maxwell’s equations with the finite-difference eigenmode (FDE) solver implemented in Ansys Lumerical MODE Solutions. We adopted a nonuniform mesh configuration to ensure high spatial accuracy while maintaining computational efficiency. The grid resolution was set to 10 nm along the propagation (x) direction, and 20 nm and 10 nm along the transverse (y) and vertical (z) directions, respectively. Within the GST region, a finer-mesh override of 2 nm and 5 nm was applied in the x and z directions, respectively, to accurately capture field variations within the PCM layer. We positioned a mode source 200 nm upstream of the switching region to excite the transverse electric (TE) mode, and placed a mode expansion monitor 200 nm downstream to record the transmitted guided power. Steep-angle perfectly matched layer (PML) boundaries were employed along the propagation axis to suppress back-reflections, whereas metallic boundary conditions were imposed in the remaining directions to confine the guided mode and prevent spurious PML amplification. This configuration ensured numerically stable simulations while avoiding impractical boundary-induced gain or loss. The total simulation duration was set to 3 ps, with an automatic field-shutoff threshold of $10^{-5}$ to ensure convergence once all transient fields had decayed. The wavelength-dependent complex refractive index data from Palik handbook \cite{palik1998handbook} were considered for Si, SiO\textsubscript{2}, and Al\textsubscript{2}O\textsubscript{3}. The complex refractive index of GST-467 is illustrated in \autoref{fig:nk_mode} and Figure S2 of the Supplementary Information. For our switching operation, we considered the third telecommunication window (C band), and the structure optimization was performed based on performance at 1550 nm. At this wavelength, silicon exhibits high transparency and low propagation losses on SOI platforms, enabling tight optical confinement and high-density integration, which are critical for compact photonic circuits \cite{keiser2011optical}.

\subsection{Optimization Metrics and Multilevel Switching Methodology}
To optimize the structure, a design figure of merit (FOM) was defined as the ratio of the extinction ratio (ER) to the insertion loss (IL). ER is the ratio of the output transmittance between the higher (amorphous) and lower (crystalline) levels, indicating the total contrast between the on and off states. Meanwhile, IL indicates the loss of light during high-state transmission, which represents the minimum loss during switching. In typical photonic switches, we want a high ER and low IL to enable low-loss, large-bit-state operations. Thus, our design FOM gives us an accurate description of the switch propagation performance \cite{zhang2024nonvolatile, meng2023electrical}, defined as follows: 

\vspace{-40pt}

\begin{align}
\label{eq:ER}
&\mathrm{ER} = 10 \log\!\left(\frac{\mathrm{T_{amor}}}{\mathrm{T_{\mathrm{cry}}}}\right) \\[-2pt]
\label{eq:IL}
&\mathrm{IL} = -10 \log(\mathrm{T_{amor}}) \\[0pt]
\label{eq:FOM}
&\mathrm{FOM} = \frac{\mathrm{ER}}{\mathrm{IL}}
\end{align}

\vspace{-10pt}
Here, T\textsubscript{amor} and T\textsubscript{cry} are the output power transmittances when PCM is in the amorphous and crystalline phases, respectively.

Furthermore, to determine the multi-level switching capabilities of our switch, the dielectric constants of intermediate crystallization levels were calculated using the effective medium theory of \autoref{eq:partial_cryst} \cite{zamani2023active, choy2015effective}. These intermediate crystallization levels result from partial crystallization, which is stable and non-volatile. 

\vspace{-15pt}
\begin{equation}
\label{eq:partial_cryst}
\frac{\varepsilon_{\mathrm{eff}}(\mathrm{P}) - 1}{\varepsilon_{\mathrm{eff}}(\mathrm{P}) + 2}
= \mathrm{P} \times \frac{\varepsilon_{\mathrm{c}} - 1}{\varepsilon_{\mathrm{c}} + 2}
+ (1 - \mathrm{P}) \times \frac{\varepsilon_{\mathrm{a}} - 1}{\varepsilon_{\mathrm{a}} + 2}
\end{equation}

Here, P denotes the intermediate crystallinity of the PCM, ranging from 0 to 1, and $\varepsilon_a$ and  $\varepsilon_c$ refer to the complex dielectric constants of the amorphous (P = 0) and crystalline (P = 1) states, respectively. The dielectric constants are calculated from the complex refractive index by following $\mathrm{\varepsilon=\left(n+ik\right)^2}$. We computed the complex refractive indices for various intermediate crystallinity levels of GST-467 using \autoref{eq:partial_cryst}, as shown in Figure S1 of the Supplementary Information.

\subsection{Multiphysics Phase Change Simulation Methodology}
A multiphysics thermo-stochastic-optical simulation framework, adapted from Wang \textit{et al.} \cite{wang2021scheme}, was employed to investigate the laser-induced transient optical responses associated with phase transitions in GST-467. The framework integrates heat transfer, stochastic crystallization dynamics, and optical field evolution within a self-consistent iterative scheme. A three-dimensional finite-difference heat conduction solver was used to calculate the transient temperature distribution resulting from Gaussian laser excitation. The resulting temperature profiles were subsequently used to drive a modified Gillespie Cellular Automata (GCA) model, which statistically simulates nucleation, growth, and dissociation processes based on classical nucleation theory and temperature-dependent viscosity. The spatially and temporally evolving crystalline fraction was then converted into effective optical constants via the Lorentz–Lorenz effective medium approximation \cite{qu2017dynamic}. Reflectance and transmittance were determined using the Fresnel equations and the characteristic matrix method, with the updated optical parameters fed back into the heat-transfer solver to ensure convergence. 
This solver analyzes the complete stack of the proposed switch, a SOI waveguide clad with the GST-467 layer. As a result, both lateral and vertical conduction into and out of these layers is captured self-consistently.
The material parameters are listed in Table S1, and the detailed computational formulation is presented in Section S5 of the Supplementary Information. Table S2 also compares the reported heat capacities and thermal conductivities of GST-467-related alloys.


\subsection{Spectroscopic Ellipsometry and Optical Model Fitting}



The complex refractive index of GST-467 was determined by spectroscopic ellipsometry using a J.A. Woollam M-2000D spectroscopic ellipsometer.
For this measurement, a 30~nm-thick GST-467 film, matching the PCM layer thickness ($\mathrm{h_{PCM}}=30$~nm) used in the optical switch simulations, was prepared by magnetron co-sputtering of elemental Ge, Sb, and Te targets at room temperature onto a silicon substrate capped with a thermally grown SiO$_2$ layer (Si/SiO$_2$), following the GST-467 deposition route described by Kusne et al. \cite{kusne2020fly}
Ellipsometric spectra were collected for both the as-deposited (amorphous) and post-annealed (crystalline) films\textcolor{black}{, the crystalline state being obtained by thermal annealing above the crystallization temperature ($\mathrm{T_c} \approx 200$--$220\,^{\circ}$C)}. 
The measured ellipsometric parameters $\Psi$ and $\Delta$ were analyzed using a multilayer optical model comprising the ambient, the film layer, and the \textcolor{black}{Si/SiO$_2$} substrate\textcolor{black}{, with the substrate optical constants taken from the literature}. 
Because spectroscopic ellipsometry with a correctly specified multilayer model returns the intrinsic, substrate-independent optical constants of the film, the resulting $n(\lambda)$ and $k(\lambda)$ are a material property of the 30~nm GST-467 layer and are therefore directly applicable to the Al$_2$O$_3$/SiO$_2$ dielectric environment of the modeled SOI device.
To retrieve wavelength-dependent optical constants, the film dispersion was parameterized using the General Oscillator (Gen-Osc) framework with a Tauc--Lorentz oscillator to describe the absorption edge and interband transitions.

\begin{figure}[!htbp]
\centering\includegraphics[width=0.96\textwidth]{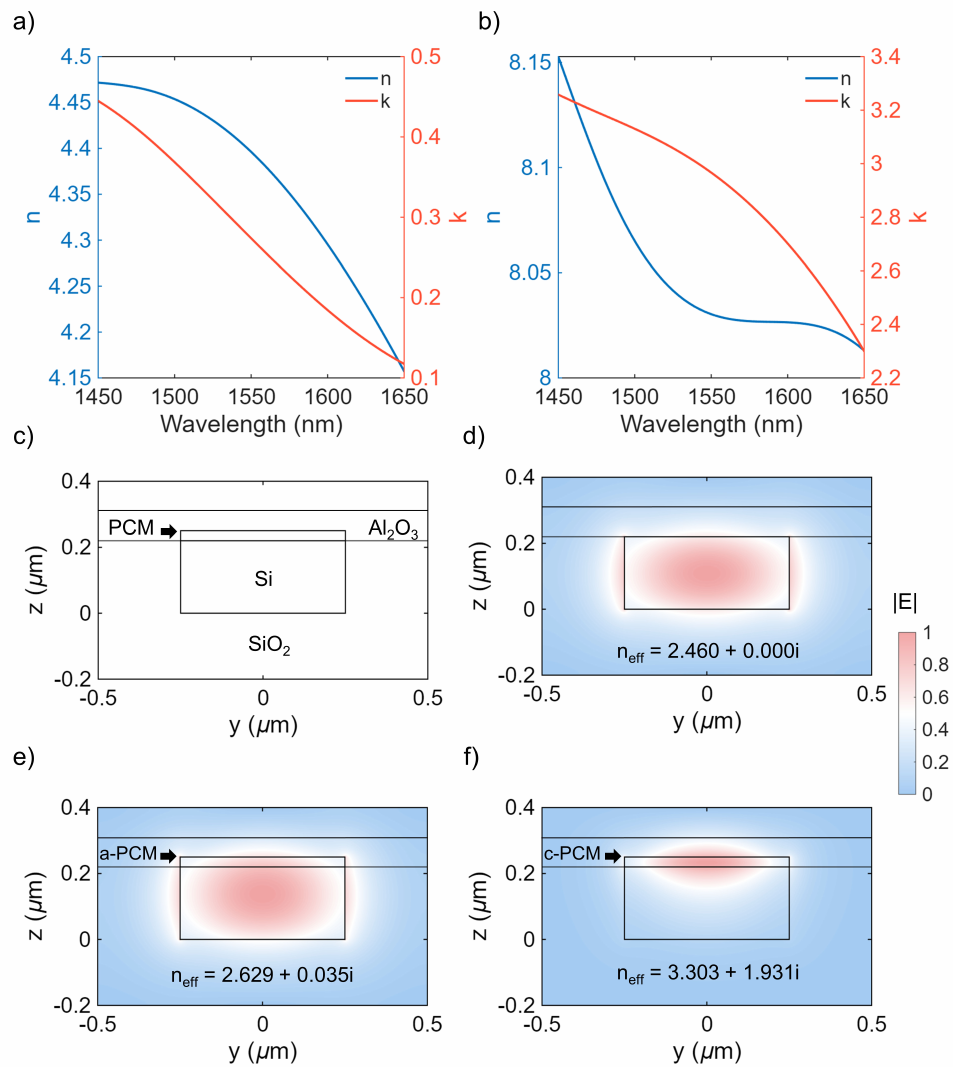}
\caption{The refractive index (n) and extinction coefficient (k) of (a) amorphous and (b) crystalline states of GST-467. (c) Cross-sectional view along the vertical y-z plane of the switch structure. Fundamental TE mode profiles of the silicon waveguide at a wavelength of 1550 nm: (d) without GST-467, (e) with amorphous GST-467 (a-GST), and (f) with crystalline GST-467 (c-GST) segments. The presence of the high-index GST cladding layer results in a lateral shift of the mode center toward the GST segment from the center of the silicon core. Furthermore, the mode becomes increasingly confined in the presence of GST, with the highest confinement observed in the c-GST case, as indicated by the elevated effective index values.}
\label{fig:nk_mode}
\end{figure}

\section{Results and Discussions}

\subsection{Ellipsometry Results}


Figures \ref{fig:nk_mode}(a) and \ref{fig:nk_mode}(b) illustrate the complex refractive indices of GST-467 in the amorphous and crystalline states, respectively. All indices show a decreasing trend with longer wavelengths. Comparing Figures \ref{fig:nk_mode}(a) and \ref{fig:nk_mode}(b), a $\sim1.8$-fold increase is observed in the real part, while a $\sim8$-fold rise is observed in the imaginary part of the refractive index. This large refractive-index contrast enabled us to design the multi-level optical switch.

\subsection{Device Physics}


\autoref{fig:nk_mode}(d-f) presents the fundamental TE mode profiles excited in the proposed 500 nm-wide, 220 nm-thick SOI waveguide structure. We chose the TE mode over the TM modes because TE modes are better confined and well-guided in the SOI platform. \cite{xu2014silicon}. \autoref{fig:nk_mode}(d) illustrates the fundamental TE mode (n\textsubscript{eff} = 2.46) of a conventional SOI waveguide, where the mode center is symmetrically aligned with the middle of the Si core. In Figures~\ref{fig:nk_mode}(e) and \ref{fig:nk_mode}(f), the mode profiles for the amorphous and crystalline PCM segments are shown, respectively. Upon introducing the GST-467 layer, the optical field becomes more tightly confined, redistributing the modal profile and enhancing field localization near the PCM/Si interface. This phenomenon is evident when comparing Figures \ref{fig:nk_mode}(e) and \ref{fig:nk_mode}(f) with \autoref{fig:nk_mode}(d).

At 1550 nm, crystalline GST-467 (c-GST) exhibits a substantially larger complex refractive index ($8.03 + 2.96i$) than crystalline Si ($3.47$). This large index leads to strong evanescent coupling of the optical mode into the PCM layer. It creates substantial overlap between the modal field and the highly absorptive PCM layer, leading to greater attenuation. In addition, the associated mode pull-up shifts the optical field away from the waveguide core, further increasing propagation loss.

In contrast, amorphous GST-467 (a-GST) has a complex refractive index ($4.39 + 0.27i$) near that of Si. This small index mismatch produces weak evanescent coupling between the light and the PCM layer, leaving the optical mode in the waveguide largely unaltered. The relatively low extinction coefficient of a-GST, together with reduced modal pull-up and slightly tighter field confinement in the core than in conventional SOI waveguides, yields higher optical transmission and lower attenuation.

Overall, the combination of higher intrinsic absorption and enhanced evanescent-field coupling in the crystalline phase results in a greater transmission disparity between the amorphous and crystalline states. This contrast enables efficient optical switching with a higher extinction ratio.

\subsection{Switch Design and Optimization}

\begin{figure}[!ht]
\centering\includegraphics[width=\textwidth]{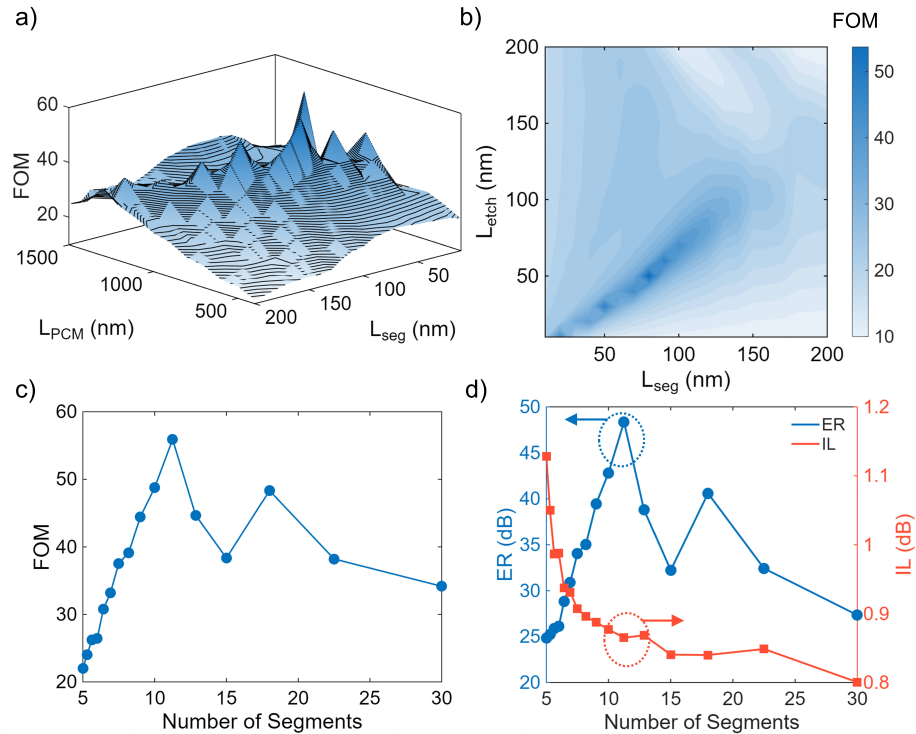}
\caption{Structure optimization results. (a) 3D surface plot of FOM as a function of $\mathrm{L_{PCM}}$ and $\mathrm{L_{seg}}$. The maximum FOM occurs at $\mathrm{L_{PCM}}$ = 880 nm and $\mathrm{L_{seg}}$ = 80 nm. (b) With $\mathrm{L_{PCM}}$ held constant at 880 nm, both $\mathrm{L_{seg}}$ and $\mathrm{L_{etch}}$ are varied to identify the optimal configuration. The highest FOM is achieved for $\mathrm{L_{seg}}$ = 80 nm and $\mathrm{L_{etch}}$ = 50 nm. Variation of (c) FOM and (d) ER and IL as a function of the number of GST segments, with fixed $\mathrm{L_{PCM}}$ and $\mathrm{L_{etch}}$ of 880 nm and 50 nm. The FOM attains its maximum at 11 segments, corresponding to $\mathrm{L_{seg}}$ = 80 nm. This maximum FOM arises from the highest ER (48.36 dB) achieved for the optimized dimensions, while maintaining a lower IL (0.86 dB).}
\label{fig:optimization}
\end{figure}

We optimized the segmented switch by varying three geometric parameters, {\color{black}$\mathrm{L_{PCM}}, \mathrm{L_{seg}}$}, and $\mathrm{L_{etch}}$, where total GST-467 segment length, {\color{black}$\mathrm{L_{PCM}}$}, is defined as:

\vspace{-12pt}



\begin{equation}
\label{eq:T_GST_v2}
{\color{black}
\mathrm{L_{PCM}}\ =\ \sum_{n} \mathrm{L_{seg,i}}
}
\end{equation}

where, n is the number of GST-467 segments and $\mathrm{L_{seg,i}}$ denotes the length of the i\textsuperscript{th} GST segment. 
The design space comprised $\mathrm{L_{seg}}$, $\mathrm{L_{etch}}\in 10-200$ nm, and $\mathrm{L_{PCM}}$ $\in 400-1000$ nm, chosen in line with prior segmented GST fabrications and standard lithographic tolerances \cite{yoon2006dry, li2016direct, tung2022nanoscale}. For each geometry, we evaluated the FOM of \autoref{eq:FOM} via a nested sweep over the specified ranges. The resulting surface in \autoref{fig:optimization}(a) exhibited a distinct maximum at $\mathrm{L_{PCM}}$ = 880 nm and $\mathrm{L_{seg}}$ = 80 nm. Holding $\mathrm{L_{PCM}}$ = 880 nm fixed, a second sweep over ($\mathrm{L_{seg}}$, $\mathrm{L_{etch}}$) identifies $\mathrm{L_{seg}}$ = 80 nm and $\mathrm{L_{etch}}$ = 50 nm as optimal, yielding FOM = 55.91. Consistent with this, \autoref{fig:optimization}(c) and (d) shows that the FOM peaks for 11 segments (i.e., $\mathrm{L_{seg}}$ = 80 nm) resulting from an ER of 48.36 dB at an IL of 0.86 dB.

Due to the segmented topology, light propagating through the structure encounters alternating high-index GST and low-index Al$_2$O$_3$ cladding regions. In the amorphous phase, the optical mode remains confined primarily within the silicon core; thus, the GST segmentation introduces only minor perturbations to the field distribution, thereby preserving a high optical transmission, T\textsubscript{amor}. However, in the crystalline state, the mode shifts toward the GST, and each GST/Al\textsubscript{2}O\textsubscript{3} interface introduces a modal mismatch that radiates power into the low-index Al\textsubscript{2}O\textsubscript{3} segments. Decreasing $\mathrm{L_{seg}}$ increases the number of segments, which enhances interfacial radiation loss in the crystalline state, while shortening the continuous guiding path. Meanwhile, $\mathrm{L_{etch}}$ governs how effectively this leaked power is dissipated before the next GST section can recouple it. Extremely small $\mathrm{L_{seg}}$ over-segments the device, adding phase-insensitive loss, whereas a very large $\mathrm{L_{seg}}$ increases IL and saturates ER. Likewise, an $\mathrm{L_{etch}}$ that is too small facilitates recoupling, thereby reducing crystalline-state loss, whereas a $\mathrm{L_{etch}}$ that is too large mainly inflates the footprint with negligible FOM benefit. The high-FOM locus in \autoref{fig:optimization}(b) follows a line of nearly constant duty cycle, preserving the phase-averaged effective index while maximizing interphase loss contrast, as confirmed by the Floquet–Bloch analysis in Section S7 of the Supplementary Information.

\subsection{Effect of the Segmentation on Performance Improvement}


To quantify the effect of segmentation, we evaluated the FOM as a function of the number of GST sections while keeping $\mathrm{L_{PCM}}$ fixed, as shown in \autoref{fig:optimization}(c). The FOM displays a pronounced maximum at 11 segments, which corresponds to $\mathrm{L_{seg}}$ = 80 nm and agrees with the two-parameter sweep in \autoref{fig:optimization}(b). Additional physical insight into this optimum is provided in \autoref{fig:optimization}(d).


For a small number of segments, $\mathrm{L_{seg}}$ is relatively large. The longer GST sections increase the light--matter interaction with a-GST, which enhances parasitic absorption in the amorphous state. Consequently, $\mathrm{T_{amor}}$ decreases, and the IL increases in the low-segmentation regime, as observed in \autoref{fig:optimization}(d). A similar increase in interaction occurs for c-GST; however, because c-GST is strongly absorbing, $\mathrm{T_{cry}}$ approaches saturation for sufficiently large $\mathrm{L_{GST}}$, indicating near-maximal attenuation in the crystalline state. Therefore, as the number of segments is increased from this low-segmentation regime, the primary improvement is the reduction of unintended amorphous-state absorption, which raises $\mathrm{T_{amor}}$ and reduces IL. This drives an increase in ER (\autoref{fig:optimization}(d)) and, correspondingly, an increase in FOM (\autoref{fig:optimization}(c)).



In contrast, at a large number of segments, $\mathrm{L_{seg}}$ becomes small. The reduced GST $\mathrm{segment}$ length weakens the interaction between a-GST and the optical mode, causing $\mathrm{T_{amor}}$ to saturate near its upper bound and IL to approach an approximately constant minimum. However, the same reduction in interaction also limits attenuation in the crystalline state: shorter c-GST sections absorb less efficiently, leading to higher $\mathrm{T_{cry}}$ with further segmentation. This degrades ER (\autoref{fig:optimization}(d)) and therefore reduces the FOM (\autoref{fig:optimization}(c)) in the highly segmented regime.

\begin{figure}[!ht]
\centering\includegraphics[width=\textwidth]{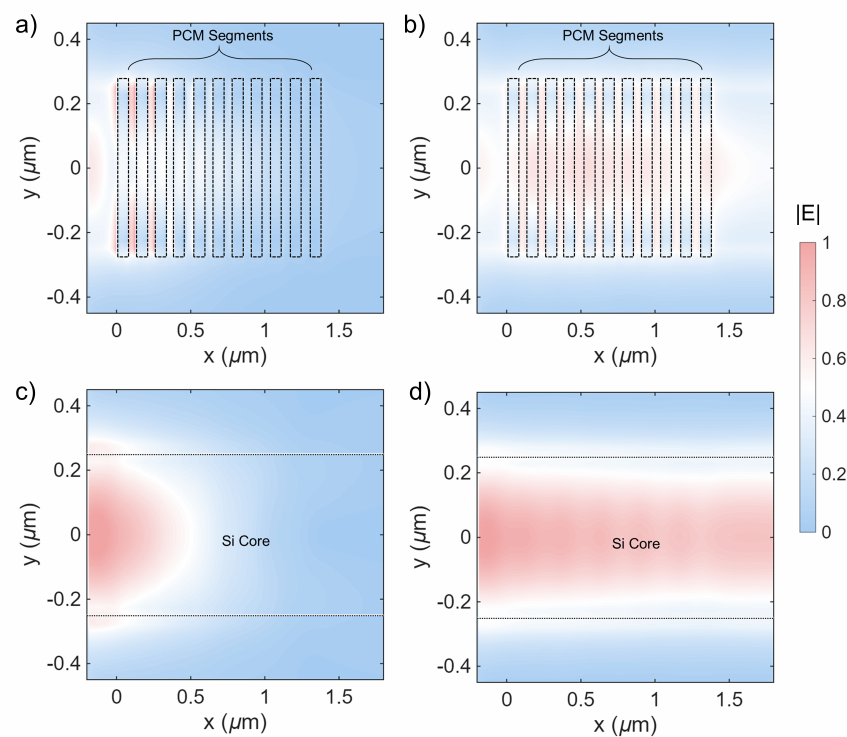}
\caption{Electric field distribution during propagation through the switch structure. Field propagation through the PCM segment is shown for (a) c-GST and (b) a-GST, while propagation through the silicon core is depicted for (c) c-GST and (d) a-GST phases. In the a-GST state, the electric field exhibits strong confinement within the silicon core, with reduced field intensity in the PCM region, indicating efficient guided mode propagation. Conversely, for the c-GST phase, the field rapidly decays along the propagation direction. Additionally, an increase in field intensity within the PCM segment is observed, signifying a lateral shift of the mode center toward the PCM layer. This shift increases absorption losses due to the high extinction coefficient of the crystalline phase, thereby contributing to field dissipation.}
\label{fig:prop}
\end{figure}

The observed optimum at 11 segments arises from the balance between these competing trends: increasing segmentation initially suppresses amorphous-state parasitic loss, whereas excessive segmentation compromises crystalline-state absorption. As a result, the maximum ER and FOM occur at 11 segments, identifying this configuration as the optimized design for our device.

Furthermore, a comparison between unsegmented and segmented implementations is summarized in \autoref{tab:segmented_vs_unsegmented}. Here, the unsegmented device has a length equal to the $\mathrm{L_{PCM}}$ of the segmented device. The optimized segmented device achieves a nearly 7-fold increase in FOM relative to the unsegmented waveguide, demonstrating the efficacy of segmentation in enhancing transmission contrast while maintaining low insertion loss.


\begin{table}[h!]
\centering
\setlength{\tabcolsep}{3pt}
\caption{Performance Comparison between Segmented and Unsegmented Structure}
\renewcommand{\arraystretch}{1.2}

\renewcommand{\tabularxcolumn}[1]{m{#1}} 

\begin{tabularx}{\textwidth}{>{\centering\arraybackslash}m{2.5cm} *{5}{>{\centering\arraybackslash}X}}
\hline
\textbf{Structure type} &
\textbf{T{\textsubscript{amor}}} &
\textbf{T{\textsubscript{cry}}} &
\textbf{ER (dB)} &
\textbf{IL (dB)} &
\textbf{FOM} \\
\hline
\textbf{Unsegmented} & 0.74 & $7.6\times10^{-2}$ & 9.91 & 1.27 & 7.75 \\
\textbf{Segmented} & 0.82 & $1.2\times10^{-5}$ & 48.36 & 0.86 & 55.91 \\
\hline
\end{tabularx}

\label{tab:segmented_vs_unsegmented}
\end{table}

Field-propagation profiles in \autoref{fig:prop} further elucidate the underlying mechanism. Figures \ref{fig:prop}(a) and \ref{fig:prop}(b) show the field evolution within the PCM section for crystalline and amorphous states, respectively, whereas Figures \ref{fig:prop}(c) and \ref{fig:prop}(d) report the corresponding fields within the silicon core. When the wave enters the etched GST section, which has a lower local index (1.63 at 1550 nm), the field becomes slightly unguided and disperses laterally, resulting in radiation loss. The effect is substantially greater in the crystalline state: as shown in \autoref{fig:nk_mode}, crystallization shifts the mode centroid toward the GST layer. It increases the complex effective index (from n\textsubscript{eff} $\approx$ 2.629+0.035i in a-GST to 3.303+1.931i in c-GST). The resulting enhancement of absorption due to modal overlap in the GST segment, together with radiative mode loss in the etched gaps, yields substantially higher loss in c-GST than in a-GST, enabling the large extinction ratio observed for the segmented design.

We further note that the segmentation does not act as a reflective grating. The segment period, $\Lambda = \mathrm{L_{seg}} + \mathrm{L_{etch}} = 130$~nm, lies well below the first-order Bragg period $\Lambda_B = \lambda_0/(2\,\mathrm{n_{eff}}) \approx 235-295$~nm, evaluated at $\lambda_0 = 1550$~nm using the real part of the modal index ($\mathrm{n_{eff}} \approx 2.629$ in a-GST and $3.303$ in c-GST). Since $\Lambda \ll \Lambda_B$, the Bragg condition is not satisfied and coherent back-reflection is suppressed by design, which is why the patterned region behaves as an effective medium (\autoref{eq:partial_cryst}) rather than as a distributed reflector, with a correspondingly low simulated device reflectance (below 1\% in both phase states, as shown in Table S3 in Supplementary Section S6).

\subsection{Switch Transmission at Different Wavelengths}



\begin{figure}[!ht]
\centering\includegraphics[width=\textwidth]{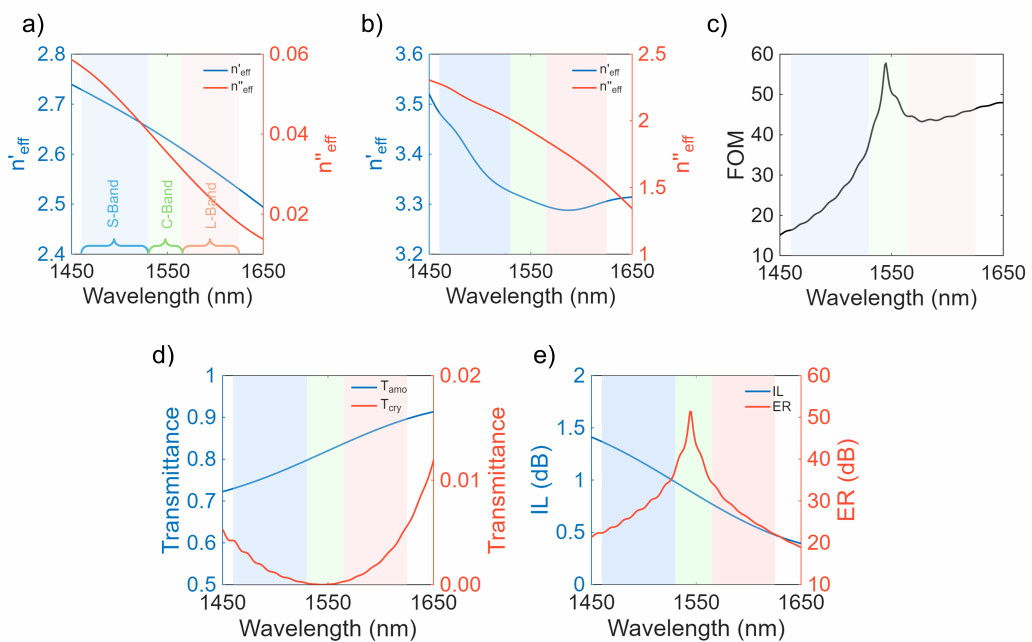}
\caption{Wavelength-dependent optical performance of the optimized segmented GST-467 switch. Real and imaginary parts of the n\textsubscript{eff} for the (a) amorphous and (b) crystalline states, respectively, showing monotonic decreases in both components with increasing wavelength. (c) Computed FOM across the S, C, and L-bands, exhibiting a pronounced peak near 1550 nm due to the optimal tradeoff between low insertion loss and high extinction ratio. (d) Simulated transmittance (unitless) spectra for the amorphous and crystalline phases, where T\textsubscript{amor} increases monotonically with wavelength while T\textsubscript{cry} reaches a minimum in the C-band. (e) Corresponding IL and ER spectra, revealing that IL decreases steadily whereas ER attains a maximum near 1550 nm, confirming the C-band as the optimal operational window for high-contrast, low-loss switching in the GST-467-based segmented photonic structure.}
\label{fig:wavelength}
\end{figure}

 Using the wavelength-dependent complex refractive indices of \autoref{fig:nk_mode}, we computed the broadband guided modes of the optimized GST-467 switch and extracted the corresponding complex effective index, $n_\mathrm{eff} = n_\mathrm{eff}' + i n_\mathrm{eff}''$. The resulting spectral response of the designed device is summarized in \autoref{fig:wavelength}, which maps the switch performance across the S, C, and L bands. Specifically, the real and imaginary components of the effective index for the amorphous and crystalline phases are shown in Figures~\ref{fig:wavelength}(a) and \ref{fig:wavelength}(b), respectively, while Figures~\ref{fig:wavelength}(c), \ref{fig:wavelength}(d), and \ref{fig:wavelength}(e) report the corresponding FOM, ER, IR, and transmission characteristics.

In the amorphous state (\autoref{fig:wavelength}(a)), both $n_\mathrm{eff}'$ and $n_\mathrm{eff}''$ decrease gradually with increasing wavelength. This trend is consistent with the monotonic reduction in the bulk optical constants of amorphous GST-467, where $n$ decreases from 4.47 to 4.16 and $k$ decreases from 0.45 to 0.1 over the same spectral window (\autoref{fig:nk_mode}(a)). The concomitant reduction of $n_\mathrm{eff}'$ and $n_\mathrm{eff}''$ at longer wavelengths slightly weakens modal confinement while reducing absorption loss, yielding an increase in $\mathrm{T_{amor}}$ (\autoref{fig:wavelength}(d)) and a corresponding decrease in IL (\autoref{fig:wavelength}(e)). 

In the crystalline state (\autoref{fig:wavelength}(b)), $n_\mathrm{eff}'$ decreases from approximately 3.55 to 3.25 with a weakly parabolic spectral dependence, whereas $n_\mathrm{eff}''$ exhibits a pronounced decline from 2.3 to 1.0. These features directly reflect the dispersion of c-GST (\autoref{fig:nk_mode}(b)): the real part of the index decreases modestly from 8.15 to 8.03 with a parabolic-like trend, while the extinction coefficient decreases more strongly from 3.2 to 2.3, indicating reduced absorption toward longer wavelengths.

The repeated c-GST/Al$_2$O$_3$ cladding topology leads to suppression of the fundamental mode in the C-band, which is evident from the reduced $\mathrm{T_{cry}}$ values in \autoref{fig:wavelength}(d). Outside this window, $\mathrm{T_{cry}}$ increases slightly in both the S-band and L-band, which in turn drives a marked reduction in ER, as shown in \autoref{fig:wavelength}(e). This spectral behavior is further reflected in the FOM plotted in \autoref{fig:wavelength}(c). The FOM peaks at 1550~nm and decreases at both shorter and longer wavelengths, closely tracking the ER trend. Notably, the degradation is more pronounced in the S-band, attributable to the higher IR relative to the L-band.



Together, these results identify the C-band as the optimal operational regime for the segmented GST-467 switch. At the same time, the persistently high FOM at longer wavelengths indicates that strong switching contrast is preserved throughout the L-band, thereby extending the usable operating spectrum beyond the C-band into the combined C and L window. This broadband, high-contrast behavior indicates robust fabrication tolerance for multilevel switching operation \cite{bogaerts2012silicon} and enables its use in matrix–vector multiplication modules in wavelength-division-multiplexing (WDM)-based photonic neural network architectures.

\subsection{Transmission window comparison with other PCMs}

\begin{figure}[!ht]
\centering\includegraphics[width=\textwidth]{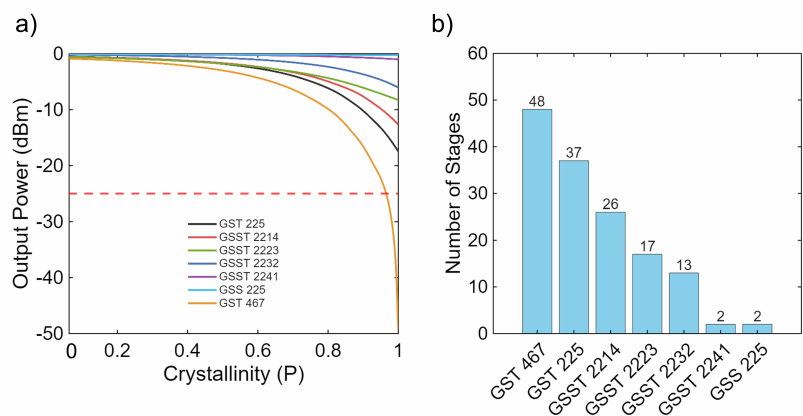}
\caption{Material-dependent multilevel response of the optimized segmented switch. (a) Output power vs. crystallinity for several PCM compositions, {\color{black} evaluated on a single, fixed device geometry, which was the optimized geometry of GST-467 switch with $\mathrm{L_{PCM}}=880$~nm, $\mathrm{L_{seg}}=80$~nm, and $\mathrm{L_{etch}}=50$~nm }(for $\mathrm{P_{in} = 0 }$ dBm). All materials exhibit a monotonic decrease in output power with increasing P, consistent with rising absorption and mode pull-up into the PCM at high crystallization states. Among these materials, {\color{black} GST-467 attains the deepest suppression, its output power falling to $\approx$ -49 dBm at full crystallization (extinction ratio $\approx$ 48 dB,  $\mathrm{P_{in} = 0 }$ dBm)}, followed by GST-225, GSST-2214, GSST-2223, GSST-2232, GSST-2241, and GSS-225. {\color{black}The dashed line marks the -25 dBm room-temperature detection floor; multilevel weights are placed only in the window above it.} (b) From the output-power characteristics, the number of resolvable levels was estimated using a conservative 0.5 dB spacing {\color{black}within the above-floor ($\geq$ -25 dBm) window}. GST-467 supports the largest state count (48 levels), with fewer levels for the other compositions{\color{black}, in a fixed-platform comparison where every material is evaluated on the GST-467-optimized geometry.}}
\label{fig:FOM_Compare}
\end{figure}



To evaluate multi-state switching performance, the complex refractive indices of the intermediate crystalline states (Figure~S1(a,b)) were applied to the fixed GST-467-optimized geometry ($\mathrm{L_{PCM}}=880$~nm, $\mathrm{L_{seg}}=80$~nm, and $\mathrm{L_{etch}}=50$~nm.), so that all materials are compared on the same patterned platform, corresponding to a shared single-mask fabrication scenario. 
For an input power of 0 dBm, the simulated output power is shown in \autoref{fig:FOM_Compare}(a). To benchmark GST-467's performance as an efficient PCM in our designed switch, the simulation model was further extended to 5 PCM compositions: Ge$_2$Sb$_2$Te$_5$ (GST-225), Ge$_2$Sb$_2$Se$_1$Te$_4$ (GSST-2214), Ge$_2$Sb$_2$Se$_2$Te$_3$ (GSST-2223), Ge$_2$Sb$_2$Se$_3$Te$_2$ (GSST-2232), Ge$_2$Sb$_2$Se$_4$Te$_1$ (GSST-2241), and Ge$_2$Sb$_2$Se$_5$ (GSS-225). The amorphous- and crystalline-state refractive indices and extinction coefficients were taken from the study by Zhang \textit{et al.} \cite{zhang2019broadband}. All materials exhibit a monotonic decrease in output power with decreasing crystallinity ($\mathrm{P}$), reflecting the concurrent increase in absorption and modal pull-up in the PCM segment as crystallization progresses. 
{\color{black} Among these, the steepest roll-off is obtained with GST-467, whose output power spans $\approx$ 0 dBm in the amorphous state to $\approx$ -49 dBm at full crystallization, almost a 48 dB swing, which is consistent with the extinction ratio and with the FDTD transmittances ($\mathrm{T_{amor}} = 0.82$, $\mathrm{T_{cry}} = 1.2\times10^{-5}$).
}
Using a conservative $0.5$~dB spacing, corresponding to about a $12\%$ change in detected power, ensures guard bands between adjacent states while remaining comfortably above the noise floor and sensitivity of integrated photodetectors. Based on this criterion, \autoref{fig:FOM_Compare}(b) summarizes the number of addressable transmission levels, consistent with the same $0.5$~dB, 5-bit (32-level) criterion used in the experimentally demonstrated, electrically programmable nonvolatile photonic memory by Chen \textit{et al.}\cite{chen2023non}. 

However, the usable level count is set not by the full $\sim48$~dB extinction of GST-467, but by the portion of the output-power window that remains above the receiver detection floor. We therefore set the levels only to $-25$~dBm, corresponding to $3.16~\mu\mathrm{W}$ at $\mathrm{P_{in}}=0$~dBm; below this limit, the signal approaches the noise-equivalent power of a room-temperature integrated InGaAs photodetector, and adjacent states are no longer reliably distinguishable \cite{chen2023non}. 

Applying these criteria allows GST-467 to support 48 levels, followed by GST-225 (37), GSST-2214 (26), GSST-2223 (17), GSST-2232 (13), GSST-2241 (2), and GSS-225 (2). Primarily, GST-467’s superior deep-extinction tail yields many more resolvable 0.5 dB states than the other PCMs.  
{\color{black}This $0.5$~dB spacing and $-25$~dBm noise floor assumption is conservative for our application since the noise-equivalent power scales with the square root of the detection bandwidth. The matrix-vector product is obtained by integrating the photocurrent over the dot-product window, rather than by operating at a per-symbol line rate. Aggregate throughput can instead be recovered through wavelength-division multiplexing and spatial parallelism, rather than increased per-lane speed \cite{feldmann2021,tait2014}. As a result, the effective per-lane detection bandwidth can remain modest, typically in the MHz to low-GHz range. This reduced bandwidth lowers the noise floor and could support additional levels below $-25$~dBm. 
Therefore, the reported counts should be regarded as a conservative lower estimate of the device's multilevel capacity.}

{\color{black}
Moreover, the decline in level count across the GSST and GSS family follows directly from the crystalline-state extinction coefficient, which falls strongly with selenium substitution (Figure~S2(d)): at 1550~nm, $\mathrm k$ decreases from about 3.1 for GST-467 to about 0.4 for GSST-2241 and 0.15 for GSS-225, so the crystalline absorption per unit length, and hence the achievable amplitude window, is intrinsically small for these low-loss compositions. 
}

{
\color{black}
Furthermore, because the discrete levels are defined in the measured output-power domain, fabrication and component variability primarily shift each device's transmission curve rather than eliminating the accessible levels. Such static device-to-device variations can be compensated with a per-device program-and-verify calibration procedure, while the conservative $0.5$~dB level spacing provides tolerance for residual deviations. This interpretation is supported by the tolerance analysis in Figure~S4, which shows that $\pm 10$~nm variations in both the segment and etch lengths, corresponding to approximately $\pm 20\%$ of the nominal $50$~nm gap, keep the insertion loss within $0.84\text{--}0.92$~dB while preserving the transmission contrast. Consequently, the number of resolvable levels is maintained across realistic fabrication process windows.
}
Collectively, these results identify GST-467 as the most suitable candidate for multilevel, high-contrast photonic switching in this architecture.

\subsection{Laser Induced Heating based Phase Change Mechanism}

GST-467 exhibits reversible transitions between amorphous and crystalline phases that can be controlled by laser heating and cooling cycles. These phase transitions underpin its non-volatile optical and memory behavior, in which the precise temporal and thermal profiles of laser excitation can control crystallization or amorphization. To explore the energy budget of both crystallization and amorphization processes, we performed multiphysics phase-change simulations under 660 nm laser heating of the PCM segments.

\begin{figure}[!hp]
\centering\includegraphics[width=\textwidth]{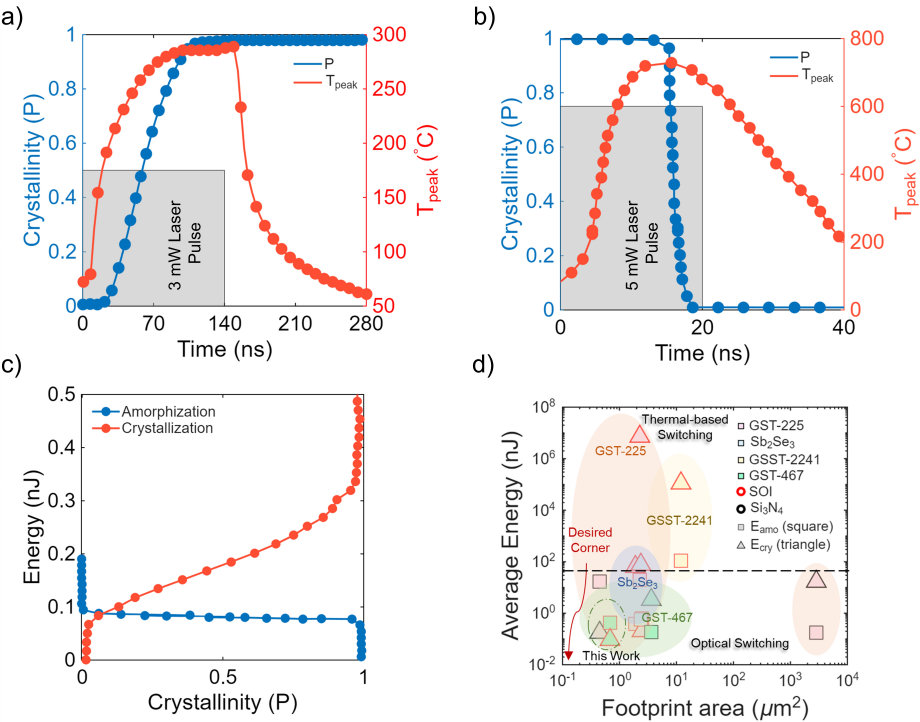}
\caption{Laser-induced phase transition dynamics and energy characteristics of GST-467 phase-change material.
(a) Temporal evolution of crystallinity (blue) and $\mathrm{T_{peak}}$ (red) under a 3 mW laser pulse, illustrating the crystallization (set) process. The $\mathrm{T_{peak}}$ surpasses 220$^\circ$C, the $\mathrm{T_{c}}$ of GST-467.
(b) Amorphization (reset) dynamics under a 5 mW laser pulse, where $\mathrm{T_{peak}}$ exceeds 540$^\circ$C, corresponding to $\mathrm{T_{m}}$. The melt-quench process causes a sharp drop in crystallinity, forming the amorphous phase (P = 0).
(c) Energy vs. crystallinity relationship for crystallization (red) and amorphization (blue). Complete crystallization requires $\sim$0.5-0.6 nJ, whereas amorphization occurs at lower energy. 
(d) Average energy versus footprint area comparison with other phase-change materials (square denotes amorphization energy, and triangle denotes crystallization energy). The dashed line separates electro-thermal (microheater) demonstrations (upper region), whose energy budget is dominated by heating the surrounding cladding and heater volume, from all-optical demonstrations (lower region), in which the absorbing PCM itself localizes the heating and enables much lower per-pulse energies. The energies reported for our design correspond to all-optical programming; on a microheater, GST-467 would be expected to require energies comparable to other PCMs of similar volume. Our design lies in the desired low-energy, small-footprint region, outperforming GST-225, $\mathrm{Sb_{2}Se_{3}}$, and GSST-2241. \cite{zheng2018gst,rude2013optical,rios2021multi,wu2018low,kato2017current,fang2022ultra}}
\label{fig:phase_change}
\end{figure}

\autoref{fig:phase_change} quantifies these phase transitions and their energy dependencies. As shown in \autoref{fig:phase_change}(a), a 3 mW laser pulse raises the peak temperature ($\mathrm{T_{peak}}$) to approximately $\mathrm{280^\circ C}$, inducing crystallization with a sigmoidal rise in crystallinity that stabilizes after cooling, confirming a non-volatile set state. In \autoref{fig:phase_change}(b), a higher-power 5 mW pulse elevates $\mathrm{T_{peak}}$ to $\mathrm{700^\circ C}$, initiating melting followed by rapid quenching, which erases structural order ($\mathrm{P}\approx0$) and restores the amorphous phase. \autoref{fig:phase_change}(c) plots the relationship between input energy and crystallinity, showing that complete crystallization requires $\mathrm{\sim0.5-0.6}$ nJ. At the same time, amorphization occurs at a lower, but abrupt, energy threshold, highlighting the reset process's low thermal budget. This indicates that, among the crystallization and amorphization processes, crystallization is easier to control because it is less abrupt. Thus, the crystallization process is named the ``Set" process, as the setup of intermediate levels is typically done during the crystallization cycle. 

{\color{black}
A specific intermediate level is therefore set by controlling a single analog quantity, the crystalline fraction $\mathrm{P}$, through the cumulative thermal dose of the programming pulses. Because the crystallization branch is gradual, the smooth, sigmoidal $\mathrm{P(t)}$ in \autoref{fig:phase_change}(a) and the monotonic energy-versus-$\mathrm{P}$ relationship in \autoref{fig:phase_change}(c) map each target level to a well-defined dose, in contrast to the abrupt amorphization branch. In practice, a level is reached deterministically by accumulative programming with a train of identical, sub-threshold pulses, as already established for on-chip phase-change synapses \cite{cheng2017chip}, combined with an optical write-and-verify step in which the device's own low-power transmission is read between pulses and the next pulse adjusted until the target level is reached. This closed-loop, single-parameter scheme can address any of the 48 levels, and, because the intermediate states are non-volatile, no holding signal is required once a level has been written.
}

A brief comparison with the other materials and platform clusters in
\autoref{fig:phase_change}(d) highlights the improvement of GST-467. Because the switching energies reported in the literature depend strongly on the actuation mechanism. Here, the dashed line in \autoref{fig:phase_change}(d) separates electro-thermal (microheater) demonstrations from all-optical demonstrations. The energy budget of electro-thermal demonstrations is primarily dominated by heating the surrounding cladding and heater volume, whereas in all-optical demonstrations, the absorbing PCM itself localizes the heating and enables much lower per-pulse energies.\cite{rios2015integrated}
First, the green-filled box with red outline represents our proposed GST-467 switch on the SOI platform, which lies closer to this desired corner,
combining a sub-$\mu$m$^2$ footprint with sub-nJ average programming energies. Using the optimized SOI geometry, the active footprint is on the order of $A\sim 0.5\times 1.38 \approx 0.69~\mu\mathrm{m}^2$, placing the device on the far-left side of the plot. In the vertical direction, both $\mathrm{E_{amor}}$ and $\mathrm{E_{cry}}$ remain in the low-energy window (sub-nJ), consistent with the energy budget extracted from the phase-transition simulations.
Second, GST-225 (orange cluster) exhibits the largest energy penalty, with crystallization energies reaching very high values (up to ~mJ in reported implementations) even at footprints of order $\sim 1$--$10~\mu$m$^2$, illustrating the well-known challenge of achieving low-energy set operations in conventional GST-based designs. These high reported values correspond to microheater-based (Joule-heating) actuation; in contrast, all-optical switching of GST-225 has been demonstrated with sub-nJ per-pulse energies, since the PCM selectively absorbs the pump light, thereby keeping the heating highly localized.\cite{rios2015integrated}
Third, GSST-2241 (yellow cluster) shifts toward larger footprints (typically $\sim 10$--$10^{2}~\mu$m$^2$) and higher energies (set energies in the $\mu$J regime), reflecting the larger thermal mass and weaker confinement typically associated with low-loss PCM platforms. 
Finally, Sb$_2$Se$_3$ (blue cluster) occupies an intermediate region: footprints remain relatively compact (order-unity $\mu$m$^2$), but the programming energies (particularly $\mathrm{E_{cry}}$) are generally tens of nJ, still separated by orders of magnitude from the desired corner.
Overall, previously reported GST-467 (green cluster) already shows a trend toward the low-energy region compared with GST-225 and GSST-2241. Still, our design advances further toward the desired corner by simultaneously reducing the footprint while maintaining sub-nJ average energies.
We note that the energies reported for our device correspond to all-optical programming; when driven by Joule heating on a microheater, GST-467 would be expected to require energies comparable to those of other PCMs of similar volume, since the energy is then governed by the heater and cladding rather than by the optical absorption of the material.
Overall, \autoref{fig:phase_change}(d) emphasizes that approaching the desired corner requires both material-level advantages (low programming energy) and geometry-level advantages (sub-$\mu$m$^2$ footprint). In this benchmark, the proposed GST-467 SOI segmented design provides one of the most favorable combined energy-area operating points among the compared PCM photonic switches.

{\color{black}
Since our laser-induced heating scheme treats the segmented switch as a collective element, we treat the 11 segments as a single thermal unit. The thermal-equilibration time across a length $\mathrm L$ is $\tau \sim \mathrm{L^{2}/D}$, with diffusivity $\mathrm{D = k/(\rho c_p)}$. For the crystalline-silicon core, $\mathrm D_{\mathrm{Si}}$ falls from about $0.9~\mathrm{cm^{2}\,s^{-1}}$ at room temperature to roughly $0.2\text{--}0.3~\mathrm{cm^{2}\,s^{-1}}$ near the melting point \cite{TouloukianVol10, GlassbrennerSlack1964}, so that, for a segment pitch $\mathrm{L_{seg}} + \mathrm{L_{etch}} = 80 + 50 = 130$~nm and a total patterned length of $1.38~\mu$m,

\begin{equation}
\tau_{\mathrm{pitch}} \approx \frac{(130~\mathrm{nm})^{2}}{\mathrm D_{\mathrm{Si}}}
\approx 0.2\text{--}0.8~\mathrm{ns}, \qquad
\tau_{\mathrm{device}} \approx \frac{(1.38~\mu\mathrm{m})^{2}}{\mathrm D_{\mathrm{Si}}}
\approx 20\text{--}95~\mathrm{ns}.
\end{equation}

Heat also crosses the 30~nm GST-467 layer into the silicon in about $0.7$~ns, using the volumetric values in Table~S1 ($\mathrm D_{\mathrm{GST}} = \mathrm{k_c/C_p} = 1.58/(1.27\times10^{6}) \approx 1.2\times10^{-6}~\mathrm{m^{2}\,s^{-1}}$). Both equilibration times are far shorter than the programming pulses of tens to hundreds of nanoseconds used in Figure~7, so adjacent segments equilibrate two to three orders of magnitude faster than the pulse ($\tau_{\mathrm{pitch}} \ll \mathrm{t}_{\mathrm{pulse}}$) and remain essentially isothermal throughout programming, crystallizing or amorphizing together, while the full patch settles to a single uniform crystallinity within the pulse. The high thermal conductivity of silicon and the sub-$\mu$m$^{2}$ footprint are therefore advantageous here, smoothing the Gaussian absorption profile and any segment-to-segment differences and making collective multilevel programming robust and repeatable.

Additionally, device-level thermal isolation is very important for our application, as we plan to implement our switch in a crossbar-style network that uses multiple switches integrated on the same platform. Thus, to quantify the thermal crosstalk between neighboring switches, we estimated the lateral thermal-diffusion length during a programming pulse:
\begin{equation}
\mathrm{L_{th}} \approx 2\sqrt{\mathrm{D_{Si}}\,\mathrm{t_{pulse}}}
\approx 1\text{--}4~\mu\mathrm{m} \quad (\mathrm{t_{pulse}} = 10\text{--}150~\mathrm{ns}),
\end{equation}
This indicates that independently addressable switches should either be separated by distances exceeding this length scale ($4~\mu\mathrm{m}$) or be provided with additional thermal isolation. In addition, the buried oxide and surrounding SiO$_2$ cladding, with a thermal conductivity of $\mathrm k_{\mathrm{SiO_2}} \approx 1.4~\mathrm{W\,m^{-1}\,K^{-1}}$, nearly two orders of magnitude lower than that of silicon, help confine heat vertically and reduce heat flow away from the active waveguide region. Furthermore, the thin 220~nm and narrow 500~nm silicon core restricts the lateral heat-conduction cross-section. If required, device-to-device thermal crosstalk can be further mitigated using established approaches, including an increased crosspoint pitch, substrate undercutting, or oxide and air isolation trenches.\cite{Jacques2019,Fang2011} 

}

\subsection{GST-467 based DNN Inference Architecture}

\begin{figure}[!ht]
\centering\includegraphics[width=\textwidth]{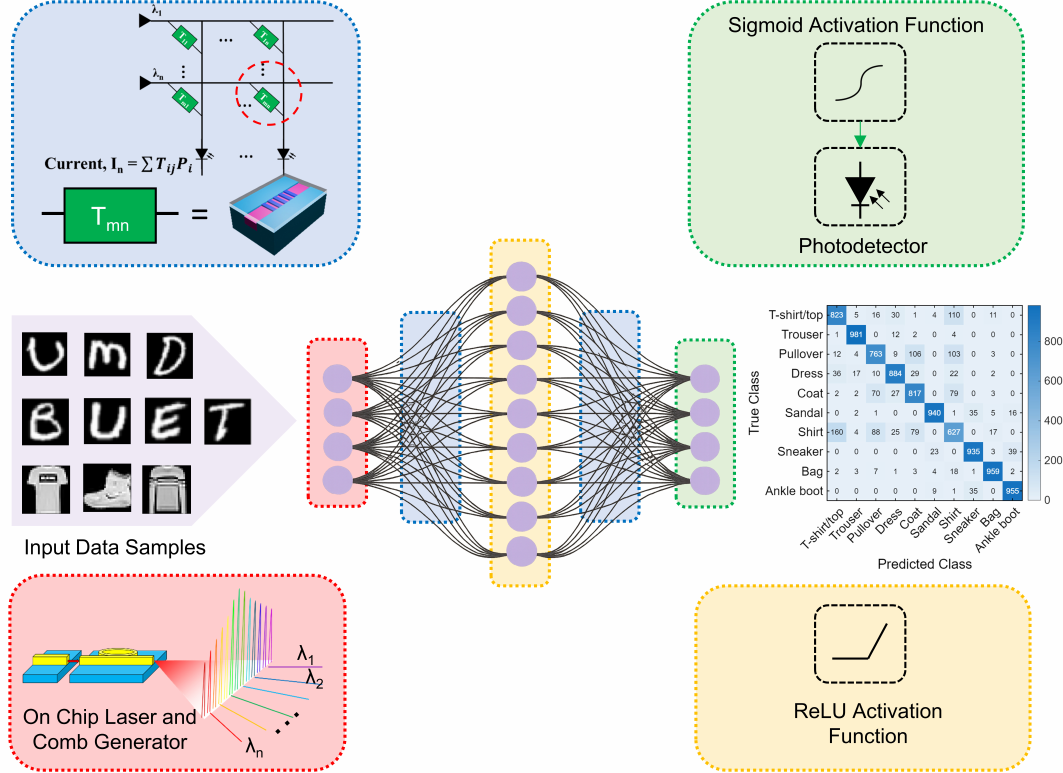}
\caption{Schematic of the photonic deep learning inference architecture.  The system comprises multiple interconnected photonic layers that execute matrix–vector multiplications and nonlinear activations. The input layer (red block) encodes optical data, which propagates through a programmable PCM GST-467-based photonic multilevel switch (blue blocks) acting as tunable weight matrices, $\mathrm{T_{mn}}$. The hidden layers (yellow block) perform reconfigurable linear transformations, while the output layer (green block) collects the processed optical signals for inference.}
\label{fig:architeture}
\end{figure}

\autoref{fig:architeture} illustrates an all-optical inference engine that uses the proposed phase-change photonic switch as a multilevel weight element. By programming intermediate crystallinity states in GST-467, each switch realizes a discrete transmission level, enabling storage of multiple weight values within a single device and supporting multilevel synaptic operation.

We adopted a simple single-hidden-layer model for benchmarking, allowing us to isolate the impact of GST-467's photonic weight. Inter-layer connections are implemented by an array of multilevel switches (blue blocks). Linear operations are performed optically, where the incident optical fields are attenuated according to the programmed transmission values, enabling matrix-vector multiplication via passive modulation of light intensity.
{\color{black}
Since the programmed crystallinity states are non-volatile, each weight is written once during an offline calibration step and then retained with zero static power, and the 660~nm beam does not modulate the weights during inference. Moreover, during inference, no continuous, real-time thermal control would be needed across the crossbar. As a result, the run-time control complexity would be independent of the array size. Thus, this scheme offers a pathway to markedly improved energy efficiency relative to electronic counterparts.
}

Parallelism is achieved by combining a waveguide crossbar topology with WDM, as shown in the blue box. Because the proposed switch exhibits a favorable figure of merit near 1550 nm, the platform is compatible with WDM-based neural inference. Practical sources for the input layer can be provided by on-chip lasers and integrated frequency-comb generators, while the compact, integrable nonlinearity blocks (ReLU and sigmoid functions) reported in prior work can supply the activation functions required for the forward pass \cite{zhou2023prospects, bai2023microcomb, huang2022programmable}. Finally, integrated photodetectors coupled to the crossbar outputs transduce the modulated optical power, proportional to the matrix-multiplication result, into electrical signals for downstream processing and readout.


{\color{black}
The array size supported by a passive crossbar is primarily determined by the optical link budget, rather than by the number of programmable levels. Along a path traversing $n$ weight cells, the maximum cascade depth can be estimated as
$\mathrm{N_{\max}} \approx (\text{link budget})/(\text{per-cell insertion loss})$.
With the sub-decibel on-state loss of the present switch ($\mathrm{IL} \approx 0.86$~dB), and assuming a representative on-chip link budget of $25-30$~dB from the integrated source to the photodetector sensitivity limit, passive cascades of approximately $30$ cells can be supported without optical gain. Larger fully connected layers can be enabled by WDM fan-out, in which each signal traverses only a limited number of weighting elements per matrix operation. Further scaling can also be supported by integrated semiconductor optical amplifiers or signal-regeneration stages.
Inter-cell back-reflection is not expected to be a limiting factor within this budget. The full-wave 3D-FDTD simulation gives a device reflectance of only $\sim 0.19\%$ in the amorphous and intermediate states (Supplementary Section S6, Table S3). Chromatic dispersion is also negligible across the micron-scale weight cells. Across the C and L bands, dispersion can be managed using standard SOI bus-waveguide engineering. A detailed system-level analysis of the link budget and noise propagation for a specified array size will be an important direction for future work.
}

\begin{table}[!ht]
\centering
\setlength{\tabcolsep}{5pt}
\caption{Performance metrics of different PCM materials for DNN inference benchmarking, with all materials evaluated on the fixed GST-467-optimized geometry}
\renewcommand{\arraystretch}{1.2}
\begin{tabularx}{\textwidth}{c *{3}{>{\centering\arraybackslash}X}}
\hline
\textbf{Material name} &
\textbf{Number of levels (0.5 dB separation)} &
\textbf{Fashion-MNIST Test Accuracy (\%)} &
\textbf{EMNIST Test Accuracy (\%)} \\
\hline
GST-467  & 48 & 87.29 & 79.90 \\
GST-225  & 37 & 87.10 & 78.27 \\
GSST-2214 & 26 & 86.84 & 77.92 \\
GSST-2223 & 17 & 85.85 & 76.82 \\
GSST-2232 & 13 & 82.36 & 69.19 \\
GSST-2241 & 2 & 25.60 & 2.18 \\
GSS-225  & 2 & 23.21 &  2.12 \\
\hline
\end{tabularx}
\label{tab:pcm_performance}
\end{table}

\subsection{Performance Comparison of GST-467 in DNN inference}

To evaluate the suitability of the GST-467 photonic switch for DNN inference, we conducted experiments on the Fashion-MNIST and EMNIST datasets. Using the calculated output power levels at an input power of 0 dBm, shown in \autoref{fig:FOM_Compare}(a), we enumerated the admissible transmission levels with a minimum power spacing of 0.5 dB. These transmission levels were then mapped to discrete synaptic weights to form the network's quantization set. The mapping procedure is detailed in Section S4 of the Supplementary Information.

With this weight set, we performed hardware-aware training to obtain final models for both datasets and then assessed classification accuracy on the held-out test sets. The network based on GST-467 achieved test accuracies of 87.29\% on Fashion-MNIST and 79.90\% on EMNIST. We further benchmarked material platforms by repeating the procedure for representative GST-, GSST-, and GSS-based switches, as shown in \autoref{tab:pcm_performance}. Across both datasets, GST-467 achieved the highest accuracy, consistent with its larger transmission dynamic range, which supports a higher effective weight precision (more distinguishable 0.5-dB spaced transmission states) and thus reduces weight-quantization error during inference \cite{kirtas2023mixed}. These results identify GST-467 as a strong phase-change platform for photonic-switch-based DNN inference.




\section{Conclusion}
In this work, we conducted a theoretical investigation of the newly emerging phase-change material GST-467 and demonstrated its strong potential for high-performance photonic switching and optical neural network applications. By experimentally extracting its optical constants and integrating them into a segmented SOI waveguide architecture, we achieved a substantially enhanced switching performance compared to conventional PCM platforms. The optimized segmented design, featuring 11 GST sections, delivered an extinction ratio of 48.36 dB, an insertion loss of 0.86 dB, and a FOM nearly 7 times that of its unsegmented counterpart. Broadband analysis further confirmed that GST-467 provides optimal contrast and minimal loss in the telecommunication C-band, with robust performance extending into the L-band.

Material-dependent comparisons established GST-467 as one of the most effective PCMs in this class. GST-467 offers the broadest transmission window, and our segmented-switch design enables higher-resolvable optical levels. The multiphysics simulations further revealed low thermal budgets for both crystallization and amorphization via laser-induced optical switching, highlighting GST-467’s suitability for energy-efficient, reversible, and stable multilevel programming.

Finally, incorporating the simulated transmission-crystallinity characteristics into photonic neural network models demonstrated that GST-467 enables superior classification accuracy on the EMNIST and Fashion-MNIST datasets compared to other PCMs. These results underscore GST-467’s unique combination of high optical contrast, multilevel stability, and low switching energy, positioning it as a promising material platform for next-generation reconfigurable photonics, non-volatile memory, and scalable in-memory photonic computing.

\begin{suppinfo}

Complex refractive index of intermediate crystallinity states of GST-467, Comparison of refractive indices among PCMs of GST, GSST, and GSS family, Proposed fabrication steps, Fabrication tolerance of the designed structure, Deep Neural Network Methodology, and Crystallization and Amorphization Workflow.

\end{suppinfo}

\begin{datastatement}
The data and code that support the findings of this study are available from the corresponding author upon reasonable request.
\end{datastatement}

\begin{acknowledgement}

The authors thank Professor Carlos Ríos Ocampo and Dr. Niloy Acharjee for valuable discussions and insights. A.S. and S.S. acknowledge the Department of Electrical and Electronic Engineering, Bangladesh University of Engineering and Technology (BUET) for providing computational facilities. 



\end{acknowledgement}


\begin{funding}
I.T. acknowledges funding provided by the National Science Foundation (awards ECCS-2210168/2210169, ECCS-2430920, and DMR-2329087/2329088), supported in part by industry partners, as specified in the Future of Semiconductors (FuSe) program.
\end{funding}

\begin{notes}
A preprint version of this work was previously made available: Sur, A.; Saha, S.; Lee, C.Y.; Takeuchi, I. Multilevel Photonic Switching in GST-467 for Deep Neural Network Inference. 2025, arXiv preprint arXiv:2512.19105. 10.48550/arXiv.2512.19105 (accessed February 14, 2026). The authors declare no competing financial interest.
\end{notes}

\bibliography{mybib}

\end{document}